\documentclass[aps,prx,superscriptaddress,nofootinbib, twocolumn]{revtex4-2}

\usepackage{amsmath,amssymb}
\usepackage{bbm}
\usepackage{graphicx}
\usepackage{booktabs}
\usepackage{bm}
\usepackage[colorlinks=true,allcolors=blue]{hyperref}
\usepackage{xcolor}

\DeclareMathOperator*{\argmax}{argmax}

\newcommand{\br}{\bm{r}}
\newcommand{\dd}{\mathrm{d}}
\newcommand{\tr}{\mathrm{Tr}}
\newcommand{\ketbra}[2]{\lvert #1 \rangle\!\langle #2 \rvert}
\newcommand{\ket}[1]{\lvert #1 \rangle}

\graphicspath{{figures/}}

\begin{document}

\title{Learned Diffractive Optics for Quantum-Optimal Inference}

\author{Matthew J. Filipovich}
\author{Alexander Duplinskii}
\author{A.I. Lvovsky}
\affiliation{Clarendon Laboratory, University of Oxford, Parks Road, Oxford, United Kingdom}

\date{\today}

\begin{abstract}
\noindent
Quantum mechanics sets the ultimate bounds on photon-limited sensing, yet practical measurements attaining these bounds are known only in special cases. 
This is particularly the case for visual sensing problems, where the goal is to infer features of a distant object based on the spatial structure of the light field it emits or reflects. 
Because of the potentially complex structure of such objects and fields, constructing optimal measurements on them is a challenging task. 
Here, we apply learned diffractive optics to state discrimination and parameter estimation of coherent and diffraction-limited incoherent light fields under a restricted photon budget.
Optimized directly on each task's figure of merit, without prior knowledge of the optimal measurement, the physically realizable diffractive optical neural networks substantially outperform standard measurements and approach the quantum limits for a given number of photons as well as in the asymptotic limit.
\end{abstract}

\maketitle
\begin{figure*}[t]
  \centering
  \includegraphics[width=\textwidth]{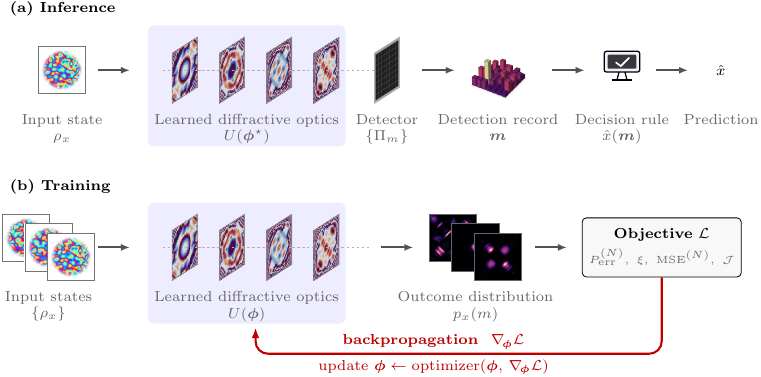}
  \caption{Learned diffractive optics for photon-limited inference.
  (a) Inference of the state $\rho_x$, encoding the unknown target $x$, using learned diffractive optical neural network with learned phase profiles $\bm{\phi}^\star$.
  The state is transformed by the DONN and subsequently detected, generating the detection record $\bm m$ of $N$ photon counts distributed among the detector bins.
  The record is mapped by the Bayes decision rule to the prediction $\hat x$.
  (b) The training procedure for optimizing the phase plate parameters $\bm{\phi}$ using a fully differentiable simulation.
  At each iteration, the input states are propagated through the DONN and the resulting output probability distributions are used to compute the objective $\mathcal{L}$, which consists of the task's figure of merit and a photon-loss penalty.
  The gradients of the objective with respect to the phase plate parameters are computed via backpropagation, and the optimizer subsequently updates the parameter values.
}
  \label{fig:overview}
\end{figure*}

\section{Introduction}

Modern computer vision typically uses one of the standard optical measurement techniques, such as direct imaging (DI), which captures an image of the object of interest, followed by \textit{in silico} post-processing.
In low-intensity scenarios, common in fluorescence microscopy and astronomy, only a finite number of photons are collected and the shot noise, arising from the discrete nature of light, becomes dominant. The performance is then set by the quantum physics of the optical measurement, which determines the task-relevant information extracted from each photon's quantum state. 
In this case, DI and other standard optical measurements are often suboptimal, losing critical quantum information carried by the field, which no conventional post-processing can help to recover.

Photon-limited inference tasks such as state discrimination and parameter estimation have known quantum limits that can often be derived analytically or numerically~\cite{personickApplicationQuantumEstimation1971, helstromQuantumDetectionEstimation1976,braunsteinStatisticalDistanceGeometry1994,jezekFindingOptimalStrategies2002}.
Measuring in the optimal basis can lead to substantial performance improvements over standard optical measurements. 
A celebrated example is estimating the separation between two diffraction-limited incoherent point sources, where measurement in the Hermite--Gaussian basis reaches the quantum limit and overcomes Rayleigh's curse, which constrains DI~\cite{tsangQuantumTheorySuperresolution2016, nairFarFieldSuperresolutionThermal2016, lupoUltimatePrecisionBound2016, paurAchievingUltimateOptical2016, pushkinaSuperresolutionLinearOptical2021, frankPassiveSuperresolutionImaging2023, rouviereUltrasensitiveSeparationEstimation2024,lvovskyPassiveOpticalSuperresolution2026}.
However, knowing the quantum limit does not automatically yield a measurement.
Moreover, even if the quantum-optimal measurement basis is known, there is no recipe for realizing it with practical optics using a limited number of optical elements with finite apertures and pixel resolution. This is especially the case when the object set of interest has a complex structure, which is common in computer vision. In this case, analytical tools for constructing the optimal measurement, which are readily available for simpler quantum sensing tasks~\cite{barnettQuantumStateDiscrimination2009}, fail.

Here we show that this challenge can be successfully tackled using the optimization tools of machine learning.  
A measurement apparatus can be optimized to approach quantum-limited performance using methods that are similar to those of training artificial neural networks, such as gradient-based inverse design. 
Specifically, we utilize diffractive optical neural networks (DONNs), which consist of a series of physically separate phase plates that apply a spatially varying phase profile to the incident wavefront, followed by an array of photodetectors at the output plane.
The plates can be realized as fabricated static elements or as programmable spatial light modulators.
DONNs have been used as multi-plane light converters, implementing unitary transformations between spatial mode bases, such as Hermite--Gaussian and Laguerre--Gaussian~\cite{morizurProgrammableUnitarySpatial2010, labroilleEfficientModeSelective2014, fontaineLaguerreGaussianModeSorter2019, kupianskyiHighDimensionalSpatialMode2023}, as well as for  machine learning tasks with classical light~\cite{linAllOpticalMachineLearning2018, wetzsteinInferenceArtificialIntelligence2020, zhouLargeScaleNeuromorphicOptoelectronic2021, rahmanUniversalLinearIntensity2023}. 

In this paper, we show that learned diffractive optics realize near quantum-optimal measurements for inference tasks under a limited photon budget, reaching a given performance with far fewer photons than standard measurements.
We demonstrate this for state discrimination and parameter estimation, at finite and asymptotic photon numbers, with coherent and diffraction-limited incoherent light.
In each case, the phase profiles are optimized directly on the task's figure of merit, such as the classification accuracy or parameter estimation precision. 
The training therefore does not require knowledge of the quantum-optimal measurement basis and automatically finds the optimal apparatus within the space of measurements the hardware can realize.

\section{Learned diffractive measurements}

The goal of each inference task is to predict a target $x$, either a discrete label or a continuous parameter, encoded in the transverse spatial degree of freedom of the input optical state $\rho_x$. The diffractive system performs custom measurements on the input photons: the DONN transforms the state, which is then detected by an array of photodetectors.
The detection record is processed using the Bayes decision rule to yield the prediction $\hat x$ [Fig.~\ref{fig:overview}(a)].

\subsection{Quantum description of the apparatus}
We assume that the field contains much less than one photon per temporal mode, allowing us to treat the field as single photons, neglecting the vacuum contribution \cite{tsangQuantumTheorySuperresolution2016,yang2017fisher}. 
The transformation of the input photon's spatial state through the diffractive optics with $L$ phase plates is given by the operator
${U = \prod_{l=1}^{L}(\mathcal{P}_l \mathcal{T}_l)}$,
where $\mathcal{P}_l$ propagates the field through free space from the $l$-th to the $(l+1)$-th plate ($\mathcal{P}_L$ to the detector) and $\mathcal{T}_l$ is the (trainable) transmission operator of the $l$-th plate.
The free-space propagation is given by
\begin{equation}\label{eq:Pl}
\mathcal{P}_l = \int_{\Omega_{l+1}} \dd^2\br \int_{\Omega_l} \dd^2\br' \, h_{z_l}(\br - \br')\,\ketbra{\br}{\br'},
\end{equation}
where $\Omega_l$ is the $l$-th aperture, $\br$ is the transverse position and the Fresnel kernel is
\begin{equation}
h_z(\bm{u}) = \frac{e^{ikz}}{i\lambda z}\,\exp\!\left(\frac{ik|\bm{u}|^2}{2z}\right),
\end{equation}
$\lambda$ is the wavelength, $z$ is the propagation distance, and $k={2\pi}/{\lambda}$. The transmission operator of the $l$-th plate is
\begin{equation}
  \mathcal{T}_l = \int_{\Omega_l} \dd^2\br\, e^{i \phi_l(\br)} \ketbra{\br}{\br},
\end{equation}
where the spatially dependent phase profile $\phi_l(\br)$ is the trainable set of parameters which defines our model.

Detection by the $m$-th detector bin corresponds to the projector \begin{equation}
  \Pi_m=\int_{\Omega_m} \dd^2\br\, \ketbra{\br}{\br},
\end{equation}
with $\Omega_m$ the bin's spatial region.
The diffractive optics and the $m$-th detector bin jointly realize a measurement on the input state defined by the operator $E_m=U^\dagger\Pi_m U$. In addition, we introduce the measurement operator $E_0=I-\sum_{m=1}^M E_m$ for the no-click outcome due to photon loss from the finite apertures of the plates and detector. The set of these operators $\{E_m\}_{m\in\mathcal{M}}$ with $\mathcal{M}\equiv\{0, 1, \dots, M\}$ constitutes a positive operator-valued measure (POVM). 
Given the input state $\rho_x$, the outcome probabilities follow Born's rule:
\begin{equation}
  p_x(m)=\tr[E_m\rho_x].
\end{equation}
Optimizing the spatially dependent phase shift in each $\mathcal{T}_l$ modifies the state transformation operator $U$ and hence the POVM. In this way, the DONN together with the detector array constitutes a trainable measurement. 

\subsection{Error functions, estimators and classifiers}
When $N$ independent photons prepared in the same state are collected by the optics, the detection record $\bm m=(m_1,\dots,m_N)$, with each $m_n\in\mathcal{M}$, is distributed as
\begin{equation}
p_x(\bm m)=\prod_{n=1}^{N} p_x(m_n).
\label{eq:factorize}
\end{equation}
The prediction is determined by the detection record using a Bayes decision rule $\hat x(\bm m)$.
For state discrimination, where the state $c$ is predicted given the prior $\pi_c$, the error probability is
\begin{equation}
  P_\mathrm{err}^{(N)}=\Big\langle\mathbbm{1}\big[\hat c(\bm m)\neq c\big]\Big\rangle_{c\sim\pi_c,\;\bm m\sim p_c},
\label{eq:perr}
\end{equation}
where $\mathbbm{1}[\cdot]$ is the indicator function. An optimal inference that minimizes this error is given by the \emph{maximum a posteriori} (MAP) rule
\begin{equation}
  \hat c(\bm m)=\argmax_{c}\ \pi_c\,p_c(\bm m).
  \label{eq:map}
\end{equation}
Combining Eqs.~\eqref{eq:perr} and \eqref{eq:map} gives rise to the objective function for training the optics:

\begin{equation}
    P_\mathrm{err}^{(N)}=1-\sum_{\bm m\in\mathcal{M}^N}\max_{1\le c\le K}\, \pi_c\,p_c(\bm m),
\label{eq:perr_multiple_states}
\end{equation}
where $K$ is the number of states to be discriminated. For binary discrimination, this simplifies to
\begin{equation}
    P_\mathrm{err}^{(N)}=\sum_{\bm m\in\mathcal{M}^N}\min  [\pi_1p_1(\bm m),\pi_2p_2(\bm m)].
\label{eq:perr_two_states}
\end{equation}

Similarly, for estimating a continuous parameter $\theta$ given the prior $\pi_\theta$, the risk is the mean-squared error
\begin{equation}
\mathrm{MSE}^{(N)}=\left\langle\big(\hat\theta(\bm m)-\theta\big)^2\right\rangle_{\theta\sim\pi_\theta,\;\bm m\sim p_\theta}
\label{eq:mse}
\end{equation}
and is minimized by the posterior mean
\begin{equation}
\hat\theta(\bm m)=\big\langle\theta\big\rangle_{\theta\sim\pi_{\theta\mid\bm m}},
\end{equation}
where $\pi_{\theta\mid\bm m}\propto\pi_\theta\,p_\theta(\bm m)$ is the posterior.

For both state discrimination and parameter estimation tasks, the optimal measurement depends on the photon number $N$ \cite{acinMultiplecopyTwostateDiscrimination2005,higgins2009mixed}. This means that the optics need to be trained specifically for each $N$. Alternatively, the training can be performed for the asymptotic case of large $N$. For binary classification, this implies maximizing the Chernoff exponent~\cite{chernoffMeasureAsymptoticEfficiency1952} 
\begin{equation}
\xi=-\ln\min_{0\le t\le1}\sum_{m\in\mathcal{M}} p_1(m)^{t}\,p_2(m)^{1-t},
\label{eq:chernoff}
\end{equation}
which characterizes the exponential decay of the error probability, $P_\mathrm{err} \sim e^{-N\xi}$.
For regression, we optimize the Fisher information per photon
\begin{equation}
\mathcal{J}(\theta)=\Big\langle\big[\partial_\theta \ln p_\theta(m)\big]^{2}\Big\rangle_{m\sim p_\theta},
\label{eq:fisher}
\end{equation} which determines the asymptotic decay of the MSE according to the Cram\'er--Rao bound, $\mathrm{MSE}^{(N)}\sim 1/[N\mathcal{J}(\theta)]$. 

In all problems solved in this paper, we have prior knowledge of the full set of input states. With this knowledge, the Bayes decision rule is optimal for each task, so the system's performance is solely determined by the realized POVM and cannot be improved by any further digital processing. This makes an important difference from traditional machine learning settings, in which the model must generalize to previously unseen instances, precluding straightforward application of the Bayes rule. 

\subsection{Training}
The TorchOptics library~\cite{filipovichTorchOpticsOpensourcePython2024}, which implements a fully differentiable model of the optical system, is used to simulate the propagation and modulation of light.
Propagation is modeled using  direct integration with the Fresnel kernel.
The physical geometry of the optical system is chosen separately for each task and is characterized by the effective Fresnel number $F=(2\sigma)^2/\lambda z$, where $\sigma$ is the width of the input beam or point-spread function and $z$ is the inter-plate distance.
Throughout this work, unless otherwise noted, we train diffractive systems with $L=8$ phase plates, each consisting of $200\times200$ superpixels; the detector also has $200\times200$ bins. Each phase plate superpixel and detector bin is simulated using $4\times4$ pixels.

The phase profiles of the DONN plates are trained using gradient-based optimization [Fig.~\ref{fig:overview}(b)] according to the task's figure of merit given by Eqs.~\eqref{eq:perr},~\eqref{eq:mse},~\eqref{eq:chernoff}, and~\eqref{eq:fisher}, which we denote as $f(\bm\phi)$.
The objective function used during training contains a one-sided quadratic term to penalize excessive photon loss ${\ell(\bm\phi)=\big\langle\tr[E_0\rho_x]\big\rangle_{x}}$:
\begin{equation}\label{eq:objective}
\mathcal{L}(\bm{\phi}) = \frac{f(\bm{\phi})}{f_{\mathrm{ref}}} + \alpha\left[\max\big(0,\,\ell(\bm{\phi})-\tau\big)\right]^2,
\end{equation}
where $f_{\mathrm{ref}}$ is the corresponding quantum bound used to normalize the first term; $\alpha$ is the photon-loss penalty weight and $\tau$ is the penalty threshold, set to 1000 and 1\%, respectively, throughout. The functions
$f(\bm{\phi})$ and $\ell(\bm{\phi})$ are both differentiable and used to calculate the gradients via backpropagation.
The Adam optimizer is used with a learning rate of 0.02 over $10^4$ iterations, which is sufficient for convergence.

We now discuss how we evaluate the figures of merit $f(\bm \phi)$. The Chernoff exponent ~\eqref{eq:chernoff}, and the Fisher information \eqref{eq:fisher} are computed by straightforward summation and integration, respectively. For the continuous-parameter MSE \eqref{eq:mse}, we sample detection records according to $\theta\sim\pi_\theta,\;\bm m\sim p_\theta$ and take the gradient from the score-function (REINFORCE) estimator~\cite{williamsSimpleStatisticalGradientFollowing1992}
\begin{equation}
\nabla_{\bm\phi}\,\mathrm{MSE}^{(N)}
=\Big\langle e^2(\bm m)\,\nabla_{\bm\phi}\ln p_\theta(\bm m)\Big\rangle_{\rm samples},
\label{eq:reinforce}
\end{equation}
where $e(\bm m)=\hat\theta(\bm m)-\theta$ is the estimation error for each sample. 

 In contrast, the binary error probability \eqref{eq:perr_two_states} can be computed without sampling. The summation over the combinatorially large set $\mathcal{M}^N$ can be obviated as follows.  We introduce the log-likelihood ratio $\Lambda(\bm m)=\ln\pi_1p_1(\bm m)/[\pi_2p_2(\bm m)]$, treating it as a scalar random variable dependent on the detection event vector $\bm m$, with the associated probability density functions $\tilde p_c(\Lambda)$. Consider the set of vectors $\mathcal{S}(\Lambda)=\{\bm m: \Lambda\le \Lambda(\bm m)\le \Lambda+\dd \Lambda\}$. Because the comparison in Eq.~\eqref{eq:perr_two_states} is determined by the ratio $\Lambda(\bm m)$, the probabilities $p_1(\bm m)$ \emph{vs.} $p_2(\bm m)$ for all elements of this set will compare in the same way, and hence so will $\tilde p_1(\Lambda)$ \emph{vs.} $\tilde p_2(\Lambda)$. Therefore Eq.~\eqref{eq:perr_two_states} can be rewritten as
 \begin{equation}
    P_\mathrm{err}^{(N)}=\int_{-\infty}^\infty\min  [\pi_1\tilde p_1(\Lambda),\pi_2\tilde p_2(\Lambda)]
    \dd \Lambda,
\label{eq:perr_two_states1}
\end{equation}
replacing the summation over the full set of possible measurement outcomes by integration over a scalar variable. 

It remains to compute the distributions $\tilde p_c(\Lambda)$. To this end, we notice that the random variable $\Lambda$ is additive: $\Lambda(\bm m)=\ln(\pi_1/\pi_2)+\sum_{n=1}^{N}\lambda(m_n)$, where $\lambda(m)=\ln[p_1(m)/p_2(m)]$ is the log-likelihood ratio associated with a single-photon event. The densities $\tilde p_c(\Lambda)$ are therefore $N$-fold self-convolutions of the corresponding single-photon distributions for $\lambda$, which can be easily computed for a given DONN parameter set $\bm \phi$. 

\subsection{{Quantum bounds}}
\label{sec:bounds}

\paragraph{Binary state discrimination.} The error probability \eqref{eq:perr_two_states} is bounded for any measurement acting jointly on all $N$ photons by the Helstrom bound~\cite{helstromQuantumDetectionEstimation1976}, which for two pure states with equal priors takes the form
\begin{equation}\label{eq:Helstrom}
P_{\mathrm{err},Q}^{(N)}=\frac12\left(1-\sqrt{1-|\langle\psi_1|\psi_2\rangle|^{2N}}\right).
\end{equation}
However, this bound is generally not achievable at $N>1$ by separable (single-copy), non-adaptive measurements --- which is the case in our study because we work in the regime of much less than one photon per temporal mode, and are limited to static interferometric state transformations. With this restriction, the bound can be derived as follows.

Consider a qubit basis $\{\ket 0,\ket 1\}$ spanning $\ket{\psi_1}$ and $\ket{\psi_2}$ such that $\ket{\psi_{1,2}}=\cos\alpha\ket 0\pm\sin\alpha\ket 1$. Suppose measurements are performed in a basis $\{\ket{v_1}=\cos\gamma \ket 0+\sin\gamma \ket 1, \ket{v_2}=-\sin\gamma \ket 0+\cos\gamma \ket 1\}$.  Then the single-photon probabilities to detect the two states in $\ket{v_1}$ are 
$q_{1,2}=\cos^2(\gamma\mp \alpha)$, and the probability of detecting $k$ of the $N$ photons in $\ket{v_1}$ is  $\binom{N}{k}q_{1,2}^{\,k}(1-q_{1,2})^{N-k}$. The $N$-copy error \eqref{eq:perr_two_states} under the MAP rule is then the sum 
\begin{equation}
P_{\mathrm{err},\mathrm{sep}}^{(N)}=\frac12\sum_{k=0}^{N}\binom{N}{k}\,\min_c\, q_c^{\,k}(1-q_c)^{N-k}.
\label{eq:perr_ind}
\end{equation}
The separable bound is this error minimized over $\gamma$, which is readily computed numerically.

The Chernoff exponent \eqref{eq:chernoff} is bounded by the quantum Chernoff exponent~\cite{audenaertDiscriminatingStatesQuantum2007}
\begin{equation}
\xi_Q=-\ln\min_{0\le t\le1}\tr\left[\rho_1^{\,t}\rho_2^{1-t}\right]\ge\xi.
\end{equation}

\paragraph{Parameter estimation.}
For collective measurements on $N$ photons, the MSE \eqref{eq:mse} is bounded by the Personick bound~\cite{personickApplicationQuantumEstimation1971,leeQuantumInspiredMultiParameterAdaptive2023}
\begin{equation}\label{eq:personick}
\mathrm{MSE}_{Q}^{(N)}=\langle\theta^2\rangle_{\pi_\theta}-\tr[\bar\rho_1 B],
\end{equation}
where  the Hermitian operator $B$ defining the optimal measurement is the solution to the equation $ B\bar\rho_0+\bar\rho_0 B=2\bar\rho_1$ with $\bar\rho_k=\int\pi_\theta\,\theta^{k}\rho_\theta^{\otimes N}\,\dd\theta$.

For separable measurements, no analytic solution is known, but the bound can be computed numerically using a trainable unitary. Specifically, we consider the full set $\{\rho_\theta\}$  of input states corresponding to possible values of the parameter $\theta$ and apply an arbitrary unitary $U=e^{X-X^\dagger}$ prior to the measurement. We compute the probability $p_\theta(m)=\mathrm{Tr}[\Pi_mU\rho_\theta U^\dagger]$ of an event in the $m$-th detector bin and subsequently the MSE \eqref{eq:mse}.
The unconstrained complex matrix $X$ is then optimized with respect to that MSE by a sampled score-function gradient \eqref{eq:reinforce}. To simplify the numerics, we use singular-value decomposition of the input state set to reduce the Hilbert space dimension.

The Fisher information \eqref{eq:fisher} is bounded by the quantum Fisher information~\cite{braunsteinStatisticalDistanceGeometry1994}
\begin{equation}\label{eq:qfi}
\mathcal{J}_Q=\tr\big[\rho_\theta L_\theta^2\big],
\end{equation}
where $L_\theta$ is the symmetric logarithmic derivative, found by solving the equation $L_\theta\rho_\theta+\rho_\theta L_\theta=2\,\partial_\theta\rho_\theta$.

\subsection{Role of coherence}
If there exists a transformation that diagonalizes  the input state $\rho_\theta$ for any value of $\theta$, the estimation problem is effectively classical: measuring in the diagonalizing basis extracts all the information present in the input state and is therefore quantum-optimal. 
Physically, such a transformation would correspond to the photon becoming fully spatially incoherent in the detector plane. An obvious example is direct imaging of natural light sources in the absence of diffraction: a natural source possesses no spatial coherence, and neither does its direct image.

This imposes an important limitation on the applicability range of our method: advantage over DI can be obtained only if the direct image has a degree of spatial coherence. Two practical cases seem to be of relevance. First, the object of interest can be inherently (partially) coherent --- for example, a laser beam or a phase-only object \cite{filipovichRoleSpatialCoherence2024,jiaPartiallyCoherentDiffractive2024}. Alternatively, coherence can be acquired during propagation --- for example, when the diffraction on the input aperture limits the imaging resolution. Although the object observed may emit fully incoherent light, propagation through the aperture imposes a degree of coherence due to the van Cittert--Zernike theorem. SPADE \cite{tsangQuantumTheorySuperresolution2016} is a classic example of the utility of an optical transformation prior to detection in this situation. 

\begin{figure}[t]
  \centering
  \includegraphics{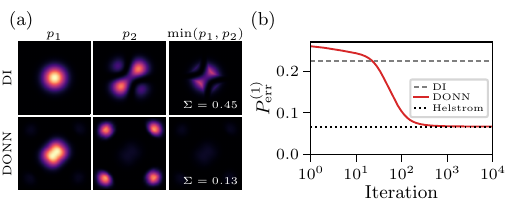}
  \caption{Diffractive optics discriminating two non-orthogonal superposition states of Hermite--Gaussian modes: $\ket{\psi_1}=\ket{\mathrm{HG}_{00}}$ and $\ket{\psi_2}=\tfrac{1}{2}\ket{\mathrm{HG}_{00}}+\tfrac{\sqrt{3}}{2}\ket{\mathrm{HG}_{11}}$ with $N=1$ photon.
  (a)~Probability distributions of the two states and their pointwise overlap in the direct image (top) and after the trained DONN (bottom).
  The summed overlap $\Sigma$ is twice the error probability \eqref{eq:perr_two_states}. 
  (b)~Error probability of the learned optics at each training iteration converging to within $0.2\%$ of the Helstrom bound.
  }
  \label{fig:disc_hg}
\end{figure}

\subsection{A simple example}
To help summarize the above information, consider the task of discriminating two non-orthogonal superpositions of Hermite--Gaussian (HG) modes: $\ket{\psi_1}=\ket{\mathrm{HG}_{00}}$ and $\ket{\psi_2}=\tfrac{1}{2}\ket{\mathrm{HG}_{00}}+\tfrac{\sqrt{3}}{2}\ket{\mathrm{HG}_{11}}$, each encoded in a single photon. The squared inner product of these states is  $|\langle\psi_1|\psi_2\rangle|^2=1/4$, corresponding to the Helstrom bound \eqref{eq:Helstrom} of $P_{\mathrm{err},Q}^{(1)}=0.067$. The minimum error achievable by direct imaging is, on the other hand, 
$P_{\mathrm{err,DI}}^{(1)}
= \frac12\int \dd^2\br \, \min\left(\left|\psi_1(\br)\right|^2,\left|\psi_2(\br)\right|^2\right) 
\approx 0.224$, making DI significantly suboptimal. 

Figure~\ref{fig:disc_hg} illustrates how a trained DONN, simulated with mode width $\sigma=20$ pixels and $F=1$, helps distinguish these states. The direct images of the two modes in the top row of Fig.~\ref{fig:disc_hg}(a) exhibit significant overlap; a photon detected in the area of this overlap does not allow one to conclusively classify its originating state. After transformation by the DONN, the overlap significantly reduces (bottom row), leading to a distinguishability error as low as $0.067$, within $0.2\%$ of the Helstrom bound.

\section{Numerical experiments}
\subsection{State discrimination}

\begin{figure}[t]
  \centering
  \includegraphics{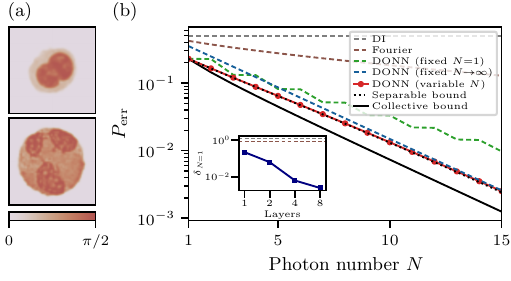}
  \caption{State discrimination of two phase objects at finite photon numbers.
  (a)~The two input states: phase profiles of an erythroblast and an eosinophil with $|\langle\psi_1|\psi_2\rangle|^2=0.70$.
  (b)~$P_\mathrm{err}$ versus photon number $N$ for the learned optics, standard measurements, and the quantum bounds.
  The dashed green and blue lines show the optics trained at $N=1$ and on the asymptotic (Chernoff) objective, respectively, evaluated across all photon numbers.
  The inset shows the relative gap $\delta_{N=1}=P_\mathrm{err}^{(1)}/P_{\mathrm{err},Q}^{(1)}-1$ of the $N=1$ learned optics versus the number of plates $L$.
  }
  \label{fig:disc_coh}
\end{figure}

Building on the above example, we first consider discrimination of two quantum states at a fixed photon number $N$. 
We train a separate diffractive optical system for each $N=1,\dots,15$ to discriminate two blood cells from the MedMNIST dataset~\cite{yangMedMNISTV2Largescale2023}, imprinted on the phase profiles of Gaussian beams [Fig.~\ref{fig:disc_coh}(a)]:
\begin{equation}
\rho_c=\ketbra{\psi_c}{\psi_c},\quad
\psi_c(\br)\propto e^{-|\br|^2/4\sigma^2}\,e^{\,i\chi_c(\br)},
\end{equation}
for $c\in\{1,2\}$, where $\sigma=25$ pixels is the beam width ($F=10$), and $\chi_c$ is the phase image of cell $c$.

The errors achieved by standard measurements and the DONN, as well as the corresponding quantum limits, are compared in Fig.~\ref{fig:disc_coh}(b). DI is blind because the intensity $|\psi_c|^2 \propto e^{-|\br|^2/2\sigma^2}$ is the same for both states. An improved performance, but still far short of the quantum limits, is shown by the far-field Fourier measurement.

The diffractive system (solid red line) approaches the separable bound (dotted black line), tracking it to within 1.4\% on average across the photon number range. The gap is attributed to hardware constraints, primarily from using a limited number of phase plates with finite apertures (see inset). Our system does not reach the ultimate quantum limit given by the $N$-copy Helstrom bound \eqref{eq:Helstrom} (solid black line) because, as discussed, reaching this bound for $N>1$ requires either a collective measurement across all $N$ photons or a series of adaptive separable measurements~\cite{acinMultiplecopyTwostateDiscrimination2005}.  Both quantum limits become equal for $N=1$.

The optics trained at $N=1$ (dashed green line) and in the asymptotic regime (dashed blue line), with the Chernoff exponent \eqref{eq:chernoff} as the objective, are also evaluated across the range of photon numbers.
As expected, each performs well in the regime it was optimized for --- near $N=1$ and at large $N$, respectively --- but
poorly outside it. The ladder-shaped behavior of the diffractive optics optimized for $N=1$ is typical for binary hypothesis testing with majority vote, see e.g.~Ref.~\cite{higginsMultiplecopyStateDiscrimination2011}.

\begin{figure}[t]
  \centering
  \includegraphics{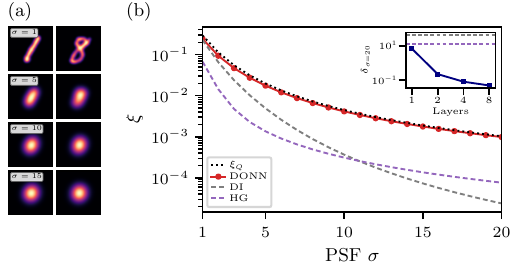}
  \caption{Diffraction-limited incoherent state discrimination in the asymptotic regime of high photon numbers.
  (a)~Blurred digits one and eight at increasing PSF width $\sigma$.
  (b)~Chernoff exponent $\xi$ versus the PSF width $\sigma$ comparing the quantum Chernoff exponent $\xi_Q$, the learned optics, DI, and the HG basis with the mode width matched to the PSF.
  The inset shows the relative gap $\delta_{\sigma=20}=\xi_Q/\xi-1$ of the learned Chernoff exponent at the largest blur $\sigma=20$ versus the number of plates $L$.}
  \label{fig:disc_incoh}
\end{figure}

An alternative coherence regime in which trained diffractive optics are beneficial involves, as discussed above, diffraction-imposed blurring.  To study this regime, we discriminate the handwritten digits one and eight~\cite{lecunGradientbasedLearningApplied1998}, blurred by a diffraction-limited Gaussian point spread function (PSF) with variable width $\sigma$ [Fig.~\ref{fig:disc_incoh}]. 
Each state is then a weighted incoherent mixture of Gaussian PSFs:
\begin{equation}
\rho_c=\sum_j w_j^{(c)}\ketbra{\psi_j^{(c)}}{\psi_j^{(c)}},\quad
\psi_j^{(c)}(\br)\propto e^{-|\br-\br_j|^2/4\sigma^2},
\end{equation}
where $\br_j$ and $w_j^{(c)}$ are the position and intensity of the $j$-th source pixel. The inter-plate distance is scaled with $\sigma$ to maintain $F=1$ across the sweep.
Because the states are significantly mixed, calculating quantum bounds for finite photon numbers is prohibitive; hence we focus on the asymptotic regime and train the optics to optimize the Chernoff exponent \eqref{eq:chernoff} across a range of PSF widths~$\sigma$ [Fig.~\ref{fig:disc_incoh}(a)].

At large blurs, the distinguishing features of the digits are sub-Rayleigh, making this a super-resolution task~\cite{graceIdentifyingObjectsQuantum2022}.
Without diffraction blurring (${\sigma=0}$), the two states are diagonal in the position basis, and DI attains the quantum limit $\xi_Q$.
At finite $\sigma$, the states are partially coherent and no longer diagonal in any common basis, rendering DI suboptimal.
Figure~\ref{fig:disc_incoh}(b) compares the Chernoff exponents achieved by the DONN, DI, and measurements in the HG basis against the quantum limit $\xi_Q$.

The learned optics approach $\xi_Q$ across the full range of blur, significantly outperforming both DI and HG: at the largest blur ($\sigma=20$), they require 41$\times$ and 13$\times$ fewer photons, respectively, to reach the same error.
Note that the HG basis is commonly used to achieve super-resolution~\cite{tsangQuantumTheorySuperresolution2016, graceIdentifyingObjectsQuantum2022, buonaiutoMachineLearningSubdiffraction2025}, but in this case performs worse than DI at low blurs.

\subsection{Parameter estimation}

\begin{figure}[t]
  \centering
  \includegraphics{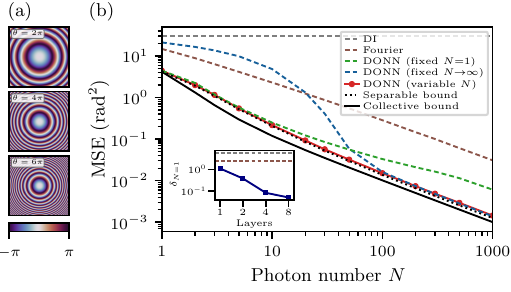}
  \caption{Estimation of axial localization at finite photon number.
  (a)~Phase profiles of the input state at various parameter values $\theta$.
  (b)~MSE versus photon number $N$ for the learned optics, standard measurements (DI and Fourier), and the quantum bounds.
  The dashed green and blue lines show the optics trained at $N=1$ and on the asymptotic objective, respectively, evaluated across all photon numbers.
  The inset shows the relative gap $\delta_{N=1}=\mathrm{MSE}^{(1)}/\mathrm{MSE}_{Q}^{(1)}-1$ of the learned $N=1$ MSE versus the number of plates $L$.}
  \label{fig:est_coh}
\end{figure}

We now turn from state discrimination to parameter estimation, where the target is a continuous parameter $\theta$ with prior $\pi_\theta$, benchmarked using the $N$-photon MSE \eqref{eq:mse}.
We consider a quadratically curved wavefront, which can be caused, for example, by defocusing the objective lens from the plane where a single point source is located. The goal is to estimate  the wavefront curvature of a Gaussian beam, quantified by the parameter $\theta$ defined as follows~[Fig.~\ref{fig:est_coh}(a)]:
\begin{equation}
\rho_\theta=\ketbra{\psi_\theta}{\psi_\theta},\quad
\psi_\theta(\br)\propto e^{-(1+i\theta)\,|\br|^2/4\sigma^2},
\end{equation}
with $\sigma=25$ pixels and $F=30$.
As previously, the optimal measurement depends on the photon budget. Since the intensity is independent of $\theta$, DI is blind.
Previous work has shown that an intensity measurement at a detection plane displaced from the focus can saturate the quantum limit at a specific parameter value in the asymptotic regime~\cite{rehacekIntensityBasedAxialLocalization2019}.
Here, we show that diffractive optics can approach this limit over a broad uniform prior $\pi_\theta$, uniform over $\theta \in [0,6\pi]$ and a wide range of photon numbers.
The DONN is trained on the exact MSE at $N=1$ and on a Monte Carlo estimate \eqref{eq:reinforce} at $N>1$.
During training, each DONN at $N>1$ is initialized from the one trained at the preceding photon number.

Figure~\ref{fig:est_coh}(b) compares the MSE as a function of $N$ for the learned optics, the quantum bounds, and standard measurements.
We observe that the diffractive optics follows the numerically estimated quantum limit for separable measurements (dotted black line), falling slightly short of the Personick limit \eqref{eq:personick} for collective measurements. The optics trained at $N=1$ (dashed green line) and on the asymptotic objective (dashed blue line), which is the prior-averaged Fisher information $\langle\mathcal{J}(\theta)\rangle_{\pi_\theta}$~\eqref{eq:fisher}, again, perform poorly outside their native regimes: the $N=1$ DONN yields $4.8\times$ the separable bound at $N=10^3$, while the asymptotic DONN exceeds the $N=1$ bound by $5.0\times$.

\begin{figure}[t]
  \centering
  \includegraphics{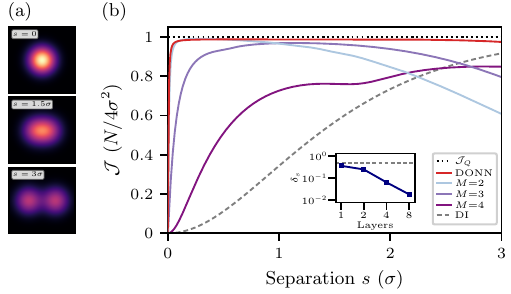}
  \caption{Estimation of the separation between two diffraction-limited incoherent point sources in the asymptotic regime.
  (a)~Two point sources at various separations $s$.
  (b)~The Fisher information $\mathcal{J}$ versus the separation $s$ from 0 to 3$\sigma$, showing the QFI $\mathcal{J}_Q$, the learned optic, DI, and mode sorters trained to discriminate the $M{\times}M$ grid of Hermite--Gaussian modes $\mathrm{HG}_{mn}$ ($m,n=0,\dots,M-1$), for $M=2,3,4$.
  The inset shows the prior-averaged relative gap $\delta_s=\langle 1-\mathcal{J}/\mathcal{J}_Q\rangle_{\pi_s}$ versus the number of plates $L$.}
  \label{fig:est_incoh}
\end{figure}

Finally, we address the well-known problem of estimating the separation $s$ between two incoherent point sources with a Gaussian PSF of width $\sigma$ [Fig.~\ref{fig:est_incoh}(a)]:
\begin{equation}
\rho_s=\tfrac12\sum_{\pm}\ketbra{\psi_\pm}{\psi_\pm},\quad
\psi_\pm(\br)\propto e^{-|\br\mp s\hat{\br}/2|^2/4\sigma^2},
\end{equation}
where the sources are displaced by $\pm s/2$ along $\hat{\br}$; here $\sigma=12$ pixels and $F=1$.
We restrict to asymptotic figures of merit for large photon numbers, focusing on the behavior as a function of the source separation. 

In this problem, the DI Fisher information \eqref{eq:fisher} vanishes as $s$ approaches zero (Rayleigh's curse).
However, the QFI \eqref{eq:qfi}  equals $\mathcal{J}_Q=1/4\sigma^{2}$ per photon, independent of $s$~\cite{tsangQuantumTheorySuperresolution2016,nairFarFieldSuperresolutionThermal2016}.
We train the DONN to maximize the average Fisher information $\langle\mathcal{J}\rangle_{\pi_s}$ over the separation prior $\pi_s$, uniform over $s\in[0,3\sigma]$.
The DONN approaches the QFI at each separation [Fig.~\ref{fig:est_incoh}(b)].
We also compare the trained DONN to multiplane light converters with the same geometry, trained to route the $M{\times}M$ grid of Hermite--Gaussian modes $\mathrm{HG}_{mn}$ ($m,n=0,\dots,M{-}1$) onto separate spatial regions of the detector by minimizing $P_\mathrm{err}^{(1)}$ \eqref{eq:perr_multiple_states}.
Averaged over the prior, the DONN reaches $98.2\%$ of $\mathcal{J}_Q$, while three HG mode sorters trained for $M=2,3,4$ attain $87.3\%$, $90.1\%$ and $68.0\%$, respectively.

We observe that directly training on the Fisher information figure of merit surpasses a sorter tailored to HG modes, even though the HG basis is known to be optimal for this problem~\cite{tsangQuantumTheorySuperresolution2016}. This apparently paradoxical behavior arises because it is impossible to train a perfect HG mode sorter with a limited number of diffractive layers, and specific imperfections can be critical for performance even if they are tiny in magnitude. For example, this particular problem requires that the detection bin for $\mathrm{HG}_{10}$ has extremely low leakage from the $\mathrm{HG}_{00}$ mode. A na\"ive mode sorter training approach is oblivious to this requirement and weights every mode equally in the figure of merit. On the other hand, with our performance-based approach, the required precision is learned automatically through training: the leakage of $\mathrm{HG}_{00}$ into the $\mathrm{HG}_{10}$ bin is only $2.3\times10^{-6}$ for the DONN, compared to $2.5\times10^{-2}$ for the $M=4$ sorter.

\section{Discussion}

These four experiments demonstrate the versatility of our method across coherent and incoherent light, finite and asymptotic photon number regimes, and discrete and continuous inference tasks.
In each case, the trained optics approach the quantum limits without prior knowledge of the optimal measurements.
Because the method requires only a differentiable model of the optics and a figure of merit that can be computed, it extends to tasks whose optimal measurement has no closed-form solution.
Even when the optimal measurement basis is known, such as with HG modes for super-resolution, our performance-optimized approach surpasses an apparatus custom-designed for that basis.
Our method can be readily implemented using spatial light modulators or printed phase masks. The optimization can additionally incorporate imperfections such as plate misalignment, yielding robust performance~\cite{menguMisalignmentResilientDiffractive2020}.

It is perhaps surprising that diffractive optics reaches the quantum-limit level of performance, even though the set of transformations attainable by these optics is only a small subset of possible transformations over the Hilbert space at hand. The dimension of that space can be estimated as the number of bins in the detector array, $200\times200=4\cdot10^4$. The space of operators over that space would have a dimension that is the square of that number, $1.6\cdot10^9$. However, the number of trainable parameters is given by the number of layers times the number of pixels in each layer: $8\times200\times200=3.2\cdot10^5$. This paradox was explored, in particular, by Kulce \textit{et al.}, who argued that the set of transformations that achieve a performance level close to optimal is quite large, and hence it is likely that an optical model with relatively few parameters can implement some elements of this set \cite{kulce2021all}. We further speculate that the \emph{effective} dimension of the  Hilbert space spanned by the input states $\rho_\theta$ with all possible values of a single parameter could be much smaller than the number quoted above. 

It is important to distinguish this work from previous research, in which optical neural networks operating at low photon numbers have been used for classification~\cite{wangOpticalNeuralNetwork2022, maQuantumlimitedStochasticOptical2025}. Those optical neural networks mimicked digital neural network architecture: optical matrix-vector multipliers were followed by single-photon detectors acting as stochastic activation functions between layers. They treated light as an incoherent intensity distribution and did not make use of the information carried in the spatial coherence of the photon. In our paper, in contrast, the diffractive optics is trained to extract this information from every photon in the best possible way allowed by quantum mechanics. As a result, our system requires fewer photons compared to a digital processor (or its optical replica) to attain the same inference accuracy, an advantage that cannot be rivaled by any amount of subsequent conventional compute. 

In our study, the input ensembles are known \emph{a priori}, and the problem complexity arises from the quantum randomness of each set of detection events. A more complex task is quantum-optimal measurement design in data-driven settings, where the DONN is trained on a limited set of states and must generalize to unseen ones during inference. In this case, the output of the quantum measurement will need to be post-processed by a digital neural network, trained in concert with the optical one. The capabilities of this approach will be demonstrated in our future research. 

Our method challenges the century-old paradigm of image analysis and integrates
quantum physics, machine learning, and advanced statistical inference into a unified framework, dramatically increasing the dimensions of Hilbert spaces on which quantum-optimal measurements can be designed and implemented. Its success on a wide range of computer vision tasks could establish a new field of \emph{quantum computational machine vision}, with the potential to redefine the limits of imaging and sensing.

\section*{Acknowledgments}
The project is funded by EPSRC Standard Grant EP/Y020596/1. 

\textit{Data availability.---} The code and data required to reproduce the results and figures reported in this article are publicly available at \href{https://github.com/MatthewFilipovich/diffractive-quantum-optimal-inference}{github.com/MatthewFilipovich/diffractive-quantum-optimal-inference}.

\bibliography{refs}

\end{document}